\documentclass[twocolumn,prl,aps,amsmath,amssymb,superscriptaddress,showpacs]{revtex4}

\usepackage{epsfig}
\usepackage{float}

\def\ami        {{\it ab--initio}\;}
\def\ai         {{\it ab initio}\;}

\def\gd         {\delta} 
\def\eps        {\epsilon}

\def\gl         {\lambda}
\def\go         {\omega}

\def\gt         {\tau}

\def\la         {\langle}
\def\ra         {\rangle}

\def\HH         {{\bf H}}

\renewcommand{\[}{\left[}
\renewcommand{\]}{\right]}
\renewcommand{\(}{\left(}
\renewcommand{\)}{\right)}

\restylefloat{figure}

\begin{document}
 
\title{{\it Ab--initio} finite temperature excitons}

\author{Andrea Marini}
\affiliation{
 Dipartimento di Fisica
 dell'Universit\`a di Roma ``Tor Vergata'',
 Via della Ricerca Scientifica, I--00133 Roma, Italy, and
 European Theoretical Spectroscopy Facility (ETSF), CNISM and
 SMC, INFM--CNR center}

\date{\today}
\begin{abstract} 
The coupling with the lattice vibrations
is shown to drastically modify the 
state--of--the--art picture of the excitonic states based on a frozen atom 
approximation.
The zero--point vibrations renormalize the bare
energies and optical strengths. Excitons acquire a non--radiative 
lifetime that decreases with increasing temperature.
The optical brightness turns out to be strongly temperature
dependent such as to induce bright to dark (and vice versa) transitions.
The finite temperature experimental optical absorption spectra of bulk Si and hexagonal BN 
are successfully explained without using any external parameter.
\end{abstract} 
\maketitle


The \ai description of the excitonic states, obtained by solving the
Bethe-Salpeter\,(BS) equation of Many--Body Perturbation Theory\,(MBPT),
constitutes a well--established approach to interpret the photoexcited properties
of bulk materials, surfaces, nanostructures and
organic/bio--molecules~\cite{rmp}.
Although 
absorption and photoluminescence\,(PL) experiments are
usually performed at room temperature, 
in the standard approach the BS equation is solved
assuming the atoms frozen in their crystallographic positions,
thus neglecting the effect of lattice vibrations.
As a consequence excitons turn out to be 
insensitive to the temperature $T$
and to have an infinite lifetime.
This is in stark contrast with the experimental results,
where the absorption and emission lines at {\it any} 
temperature 
show an intrinsic width that reflects the
{\it finite} lifetime of the underlying excitonic states.
Moreover, in bulk  semiconductors, 
it is a well known fact
that the absorption line position, width,
and intensity 
show a clear $T$ dependence~\cite{cardona_ssc}. 
In the frozen--atom BS equation this dependence is not described at all.
Even in the  $T\rightarrow\,0$ limit, where
atoms vibrate to fulfill the uncertainty principle (zero--point vibrations),
the calculated absorption spectra is
commonly convoluted with some artificial,
{\it ad--hoc} numerical broadening function chosen to yield
the best agreement with the experiment.
More generally the finite temperature non--radiative damping, and the energy and 
optical strength renormalization, define the quantum efficiency
of the excitons as photo--emitters, a key parameter in devising materials 
for optoelectronic applications.

Bulk silicon\,(Si) and hexagonal boron nitride\,({\it h}--BN) are two 
paradigmatic semiconductors 
whose optical properties show remarkable differences.
Si is one of the most deeply investigated material in the \ai community~\cite{rmp}.
Although its optical properties have been studied in different 
theoretical frameworks~\cite{rmp}
the finite temperature dielectric function,
measured by
Lautenschlager and Jellison~\cite{si_exp} twenty years ago,
remains still unexplained.
While Si has a small indirect gap, 
{\it h}--BN is 
a wide direct gap quasi--two--dimensional semiconductor~\cite{hBN_dft}.
At difference with Si, {\it h}--BN  optical spectra are dominated by
a bound exciton with large binding energy ($\sim$0.7\,eV).
Very recent results by Watanabe et al.~\cite{watanabe} stimulated
interest in this material for the possible applications as ultraviolet 
laser device.
In the same experiment Watanabe observes a  
rich series of PL peaks that appear together with the free exciton
bands pointing to a non--negligible effect of the lattice vibrations.

In this Letter I solve, in a fully \ami manner,
the Bethe--Salpeter equation including the coupling with the lattice vibrations.
The picture of the excitons obtained within a frozen--atom approximation turns out to 
be deeply modified,
both at zero and finite temperature.
Excitons acquire a non--radiative lifetime, otherwise infinite in the frozen--atom approximation.
The finite temperature optical spectra of Si and {\it h}--BN 
are reproduced in excellent agreement with the experimental results.
The thermal properties of the excitonic states
are explained in terms of a weak (Si) and a strong ({\it h}--BN)
exciton--phonon coupling.
In Si the lattice vibrations affect only the electron--hole
substrate of the excitonic states,  while in  {\it h}--BN,  
they participate actively in
the excitons build--up.
In {\it h}--BN, this strong coupling  induces bright to dark (and vice versa) transitions
and reduces, at zero temperature,
the lowest exciton binding energy by $\sim$30\%.

In the frozen--atom\,(FA) BS equation 
the excitonic states $|\gl_{FA}\ra$ and energies $E^{FA}_{\gl}$ are
eigenstates and eigenvalues of the Hamiltonian $\HH^{FA}$, written in 
the electron\,(e) hole\,(h)  basis~\cite{rmp}
\begin{align}
  H_{\substack{ee'\\hh'}}^{FA}=\(E_e -E_h\)\gd_{eh,e'h'}+
  \(f_e-f_h\)\Xi_{\substack{ee'\\hh'}},
  \label{eq:1}
\end{align}
with $E_{e\(h\)}$ and $f_{e\(h\)}$ the quasi--electron (hole) energies and 
occupations. $\Xi$ is the Bethe--Salpeter kernel,
that is a sum of a direct and an exchange electron--hole\,(e--h)
scattering: $\Xi_{\substack{ee'\\hh'}}=\la e h| W-2V|e'h'\ra$.
$W$ is the statically screened and $V$ is the bare Coulomb interaction.
The absorption spectrum is given by the imaginary part of the dielectric
function, $\eps_2\(\go\)=8\pi/V \sum_{\gl}|S^{FA}_{\gl}|^2\Im\[\(\go-E^{FA}_{\gl}+i\eta\)^{-1}\]$
where $S^{FA}_{\gl}=\la GS|i\hat{\xi}\cdot\vec{r}|\gl_{FA}\ra$ are the excitonic optical
strengths,
$\eta$ is a broadening parameter, $V$ is the crystal volume,
and $\hat{\xi}$ the light polarization direction.
Eq.(\ref{eq:1}) is \ai because the single--particle quasi--energies and the
kernel $\Xi$ are calculated starting from 
Density--Functional Theory\,(DFT) wavefunctions and energies, 
with no adjustable parameters~\cite{calc_details}.
In the standard approach, the quasiparticle\,(QP) energies $E_{e,h}$,
obtained within the GW approximation for the
electronic self--energy~\cite{rmp},
are assumed to
be real and independent on $T$. 
This approximation is justified by the fact that the smallest 
excitation energy in a semiconductor, the gap energy $E_g$, 
is usually much larger than the thermal energy corresponding
to typical experimental temperatures, i.e. $T\ll E_g/k_b$.
The Hamiltonian $\HH^{FA}$ is then hermitian~\cite{rmp} and 
$T$ independent. As a consequence, 
the energies $E^{FA}_{\gl}$ are real, and $\eta$ is used as an {\it a posteriori}
parameter to mimic the experimental broadening of the absorption 
peaks~\cite{albrecht_cardona_comment}. 

In the finite temperature regime the levels $E_i$ 
acquire an explicit dependence on the temperature,
$E_{i}\(T\)=E_i+\Delta E_i\(T\)$, with 
$\Delta E_i\(T\)=\Delta E^{e-p}_{i}\(T\)+\Delta E^{TE}_i\(T\)$.
$\Delta E^{TE}$ is the thermal expansion\,(TE) contribution~\cite{cardona_ssc,thermal_expansion}.
$\Delta E^{e-p}$ 
represents the {\it complex} energy correction
that arises from the electron--phonon\,(e--p) interaction.
In this work the e--p interaction is treated in the Heine, Allen, and Cardona
approach~\cite{cardona_ssc} where
$\Delta E^{e-p}$ can be rewritten
in terms of an e--p coupling function $g^2F_i\(\go\)$,
\begin{align}
  \Delta E^{e-p}_{i}\(T\)=\int d\go\, g^2F_i\(\go\)\[N\(\go,T\)+1/2\],
  \label{eq:2}
\end{align}
with 
$N\(\go,T\)=\(e^{\beta \go}-1\)^{-1}$ being the Bose occupation function.
The complex $g^2F_i$ function is given by
\begin{align}
  g^2F_i\(\go\)=\sum_{\nu} \frac{\partial E_i}
  {\partial N\(\go_{\nu},T\)}\gd\(\go-\go_{\nu}\),
  \label{eq:3}
\end{align}
with the sum is extended to all phonon modes $\nu$~\cite{dfpt_calcs}.
Eqs.(\ref{eq:2}--\ref{eq:3}) tells that,
$\Re\[\Delta E^{e-p}_i\]$ arises from the quadratic contribution to
the expansion of $E_{i}\(T\)-E_i$ in the atomic displacements.
As shown in Ref.~\cite{Lautenschlager},
the QP states can also decay emitting phonons, thus acquiring a finite lifetime
($\propto 1/\Im\[\Delta E^{e-p}_i\]$).
Phonons have very small energies ($\approx\,$100\,meV) and,
consequently, can be populated at the typical experimental temperatures.

The temperature dependence of the QP states, arising from the
e--p interaction, modifies Eq.(\ref{eq:1})~\cite{kernel_correction}. 
As the $E_{e,h}\(T\)$  functions are 
complex, the BS Hamiltonian turns in a 
{\it non hermitian} operator
\begin{align}
  H_{\substack{ee'\\hh'}}\(T\)= 
  H_{\substack{ee'\\hh'}}^{FA}+\[\Delta E_e\(T\) -\Delta E_h\(T\)\]\gd_{eh,e'h'},
  \label{eq:4}
\end{align}
and the excitonic states are solution of the eigenproblem 
$H\(T\)|\gl\(T\)\ra=E_{\gl}\(T\)|\gl\(T\)\ra$.
The eigenstates  $|\gl\(T\)\ra$ are 
linear combinations of e--h pairs:
$|\gl\(T\)\ra=\sum_{eh}A^{\gl}_{eh}\(T\) |eh\ra$,
with $A^{\gl}_{eh}=\la eh|\gl\ra$. 
If we plug this expansion in the definition of 
the excitonic energies 
$E_{\gl}\(T\)=\la\gl\(T\)|\HH|\gl\(T\)\ra$ we get
\begin{multline}
  E_{\gl}\(T\)=\la\gl\(T\)|\HH^{FA}|\gl\(T\)\ra +\\
  \sum_{eh} |A^{\gl}_{eh}\(T\) |^2\[\Delta E_e\(T\) -\Delta E_h\(T\)\].
  \label{eq:5}
\end{multline}
Using Eq.(\ref{eq:2}) and neglecting the 
TE term, Eq.(\ref{eq:5}) yields
\begin{multline}
  \Re\[\Delta E_{\gl}\(T\)\]=
  \[\la\gl\(T\)|\HH^{FA}|\gl\(T\)\ra - \la\gl_{FA}|\HH^{FA}|\gl_{FA}\ra\]+\\
  \int d\go\, \Re\[g^2F_{\gl}\(\go,T\)\]\[N\(\go,T\)+1/2\],
  \label{eq:6}
\end{multline}
\begin{align}
  \Im\[E_{\gl}\(T\)\]=
  \int d\go\, \Im\[g^2F_{\gl}\(\go,T\)\]\[N\(\go,T\)+1/2\],
  \label{eq:7}
\end{align}
where $\Delta E_{\gl}\(T\)=  E_{\gl}\(T\)-E^{FA}_{\gl}$ and
I have introduced the exciton--phonon coupling function
$g^2F_{\gl}\(\go,T\)=\sum_{eh} 
|A^{\gl}_{eh}\(T\)|^2 \[g^2F_e\(\go\)-g^2F_h\(\go\)\]$.
Eqs.(\ref{eq:6}--\ref{eq:7}) constitute a key result of this work.
Eq.(\ref{eq:7}) defines, in an \ami manner, the non--radiative excitonic lifetime
$\gt_{nr}^{\gl}=\[2\Im\(E_{\gl}\(T\)\)\]^{-1}$, that is otherwise
infinite in the FA approximation.
The dielectric function now depends explicitly on $T$, 
$\eps_2\(\go,T\)=8\pi/V \sum_{\gl}|S_{\gl}\(T\)|^2\Im\[\(\go-E_{\gl}\(T\)\)^{-1}\]$
and no damping parameter $\eta$ is needed anymore.

Eq.(\ref{eq:6}) defines the temperature dependence of the excitonic
energies and is composed of two contributions:
the integral of the $g^2F_{\gl}$ function
arises from the 
renormalization of the
electron--hole pairs $|eh\ra$ that constitute the excitonic packet (with amplitudes
$A^{\gl}_{eh}$).
This term represents 
an {\it incoherent contribution}, where the electrons and holes
interact separately with the lattice vibrations.
The first term, instead, describes an active participation 
of the phonon modes in the excitonic state build--up. It is, then,
a {\it coherent} contribution that
modifies the $A^{\gl}_{eh}$ components, and
vanishes when $|\gl\(T\)\ra=|\gl_{FA}\ra$.
Thus  Eqs.(\ref{eq:6}--\ref{eq:7})  define two physical regimes of the
exciton--phonon interaction: in the {\it weak coupling} case  $|\gl\(T\)\ra\approx|\gl_{FA}\ra$,
and the incoherent contribution
is dominant 
(this is the case of Si).
{\it h}--BN, instead, belongs to 
the {\it strong coupling} case, where the coherent term in  Eq.(\ref{eq:6}) cannot be
neglected.

A remarkable property of  Eqs.(\ref{eq:6}--\ref{eq:7})  is that although
$N\(\go,T\rightarrow 0\)=0$, the excitonic energies do not reduce to the FA
values  and the excitonic lifetimes remain finite when $T\rightarrow 0$, 
because of the $1/2$ factor.
This factor arises from the quantum--mechanical vibrations of the
atoms when $T=0$ (the so--called zero--point vibrations~\cite{born}).

\begin{figure}[H]
\begin{center}
\epsfig{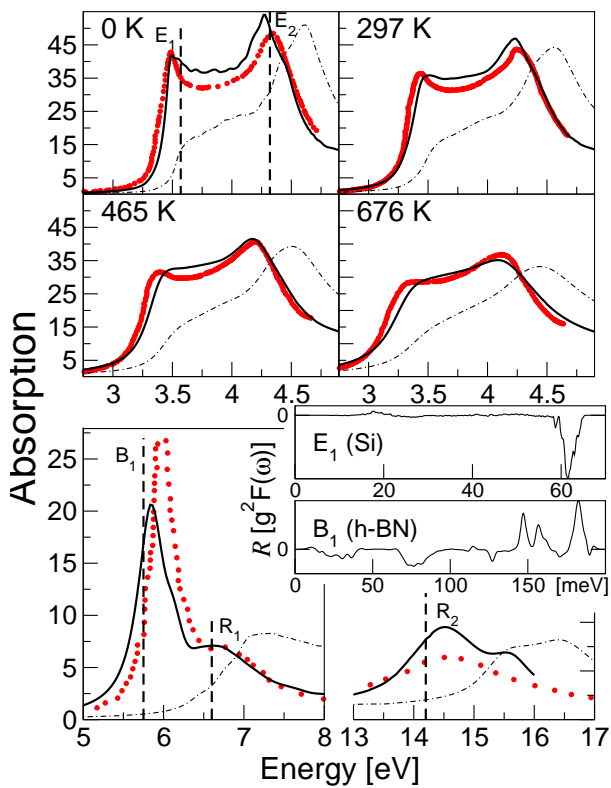}
\end{center}
\vspace{-.5cm}
\caption{\footnotesize{
(color on--line)
Optical absorption of bulk Si (upper frames) for several temperatures and 
of {\it h}--BN (lower frames) at room temperature. 
The experimental spectra~\cite{si_exp,hbn_exp} (circles) are compared with the 
BS equation (solid line), and with the independent particle approximation (dot--dashed line).
In the insets the exciton--phonon spectral functions $\Re\(g^2F_{\gl}\(\go,T=0\)\)$ are shown for the
$E_1$ (Si) and $B_1$ ({\it h}--BN) peaks (see text).
The width of the absorption peaks reflect the damping of the excitons
due to the scattering  with phonons. No additional numerical damping is included.
The excitonic energies obtained within the frozen--atom BS equation 
(represented by the vertical dashed lines)
are red--shifted in Si and blue--shifted in {\it h}--BN. 
}}
\label{fig:1}
\end{figure}

The experimental finite temperature optical spectra of Si~\cite{si_exp}, shown in
Fig.(\ref{fig:1}), are dominated by two excitonic peaks (E$_1$ and E$_2$) resonant
with the electron--hole continuum and characterized by a moderate e--h attraction.
As the temperature increases,
the E$_{1,2}$ peaks move towards lower energies,
with a width that increases with $T$. 
This gradual red--shift 
has been
studied only in an independent--particle approximation\,(IPA)~\cite{Lautenschlager}, thus neglecting
excitonic effects.
While the IPA shows only a weak dependence on $T$ 
both peak position and widths of the  E$_{1,2}$ peaks  are well reproduced by 
the results of the finite--$T$ BS equation, shown in Fig.(\ref{fig:1}).
Excitons acquire a finite damping, that starting from $\sim 30$\,meV at
$T=0$\,K and increasing to $\sim 60$\,meV at room--temperature and 
$\sim 150$\,meV at $T=676$\,K, is in excellent agreement 
with the experimental estimations~\cite{Lautenschlager}.
Compared with the frozen--atom BS equation the position of the E$_1$ and E$_2$ peaks
at $T=0$ is red--shifted by $80$\,meV,
so to correct the deviation of previous calculations~\cite{albrecht_cardona_comment}
from the experimental spectrum.

The $g^2F_{\gl}$ function can be now used 
to pin down the phonon modes that contribute to the red--shift of the 
E$_{1,2}$ peaks.
In the inset of Fig.(\ref{fig:1}) the $\Re\(g^2F_{\gl}\(\go\)\)$ for the E$_1$ state
shows that the exciton is mainly coupled with the optical phonons (60\,meV 
peak), with the acoustic branches giving only a small correction.
As the temperature increases, 
the phonon population $N$ in Eqs.(\ref{eq:6}--\ref{eq:7}) also increases, thus enhancing the red--shift
and the width of the optical peaks, and leading to a linear scaling with the
temperature when $T\ge 200$\,K and $N\(\go,T\)\sim1/\beta T$.
A more careful analysis of the different contributions to $\Delta E_{\gl}\(T\)$
given by Eq.(\ref{eq:6}) 
shows that the incoherent contribution (second term in r.h.s. of Eq.(\ref{eq:6})) is dominant. 
This is due to the fact that
the moderate e--h attraction prevents the E$_{1,2}$ excitons to behave 
as a unique, bosonic--like, particles.
Consequently the lattice vibrations mainly couple with the e--h substrate of the
excitons.
It is important to note that in this case, Eqs.(\ref{eq:6}--\ref{eq:7}) 
can be simplified using the result of the FA BS equation, as
$\Delta E_{\gl}\(T\)\approx \int d\go\, \Im\[g^2F_{\gl}\(\go\)\]\[N\(\go,T\)+1/2\]$,
with $|\gl\(T\)\ra\approx|\gl_{FA}\ra$.
\begin{figure}[H]
\begin{center}
\epsfig{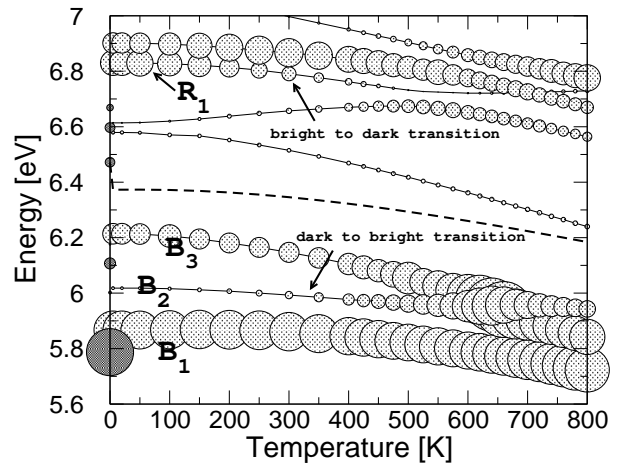}
\end{center}
\vspace{-.5cm}
\caption{\footnotesize{
Temperature dependence of the energies and oscillator strengths
of the near--gap
excitons (bound and resonant) in {\it h}--BN. The dashed line indicates the energy position of
the optical gap in the independent--particle approximation.
The size of the circles are proportional to the excitonic
optical strength. The darker circles at T=0 is the result obtained
neglecting exciton--phonon coupling.
At T=0 we observe a $30\%$ reduction of the $B_1$ exciton binding energy due to
the zero--point lattice vibrations. The $B_2$ and $R_1$ excitons, instead, undergo
a bright to dark (and vice versa) transition at room-temperature (see text).
}}
\label{fig:2}
\end{figure}

{\it h}--BN is a anisotropic, insulating compound, 
consisting of graphite--like sheets with an hexagonal structure 
arranged in an ABAB$\dots$ stacking~\cite{hBN_dft}.
The optical and electronic properties, as well as 
the lattice dynamics~\cite{serrano} are strongly influenced by  the layered
structure.
The in--plane experimental optical absorption spectrum measured at
room--temperature~\cite{hbn_exp} is shown in Fig.(\ref{fig:1}), lower frames.
Three prominent peaks are clearly distinguishable: a bound state
B$_1$ at $5.98$\,eV, and two resonant states, R$_1$ at $6.87$\,eV and
R$_2$ at $14.7$\,eV.
The frozen--atom BS equation predicts the three peak energies to be  $5.75$\,eV,
$6.6$\,eV, and $14.2$\,eV~\cite{arnaud},
and $0.1-0.5$\,eV red--shifted if compared to the experiment.

The room--temperature solution of the BS equation is compared with the experiment
in Fig.(\ref{fig:1}).
Both experimental peak positions and widths are well described,
and the B$_1$, R$_1$, and R$_2$ states are blue--shifted of $0.07$\,eV,
$0.17$\,eV, and $0.3$\,eV compared to the frozen--atom BS equation results. 
The different sign of the phonon induced corrections of the excitonic peak
positions is the first striking difference with the case of Si and
can be understood looking at the function 
$\Re\(g^2 F_{\gl}\(\go\)\)$ for the B$_1$ state, showed in 
the inset of Fig.(\ref{fig:1}).
The  anisotropic structure of  {\it h}--BN  is reflected in
the rich series of phonon peaks in the $g^2 F_{\gl}$ function.
As showed by Serrano et al.~\cite{serrano},
the phonon modes corresponding to the peaks at $\sim$30\,meV and $\sim$75\,meV
are polarized 
perpendicularly to the
hexagonal layers. 
As the bound excitons are spatially confined within the 
layer~\cite{arnaud}, these modes tend to stretch the layers thus
increasing the exciton localization,
and consequently, its binding energy.
The high--energy modes ($\go\geqslant\,100\,meV$), instead, 
are polarized parallel to the layer. These modes correspond to in--plane
vibrations that interfere  with the binding of the e--h pairs embodied in the
excitonic state, counteracting the excitonic localization.
Their stronger positive contribution to the $g^2 F_{\gl}$ function 
causes 
an overall blue--shift of the absorption peaks, 
and  a reduction of the exciton binding energy.
Similarly to the case of Si, the {\it h}--BN
QP optical gap 
is shrank by the electron--phonon coupling by $0.12$\,eV.
Thus we get an overall reduction of the lowest
exciton binding energy of $0.2$\,eV, that is $30$\,\% the value obtained
neglecting the exciton--phonon coupling ($0.72$\,eV).

The thermal evolution of the excitonic energies and optical strengths $|S_{\gl}\(T\)|^2$
for the near--gap excitons is showed in Fig.(\ref{fig:2}). 
The size of the circles is proportional to $|S_{\gl}|^2$.
The opposite contribution to the $g^2 F_{\gl}$ function
of the low and high--energy phonons makes the 
excitonic energies 
to be almost constant for $T\leq 500$\,K,
in agreement with the experimental observation~\cite{silly}.
In contrast, the excitonic optical strength
drastically depends on the temperature.
We see that the R$_1$ (resonant) 
and the B$_2$ (bound) 
excitons undergo a bright to dark (and vice versa)
transition at room--temperature.
Indeed, we have that $S_{\gl}\(T\)-S^{FA}_{\gl}=\la GS|e^{i\hat{\xi}\cdot\vec{r}}
\[|\gl\(T\)\ra-|\gl_{FA}\ra\]$, so that 
this astonishing effect is entirely due to the coherent contribution
$\Delta\[E_{\gl}\(T\)\]$, given by the first term in 
Eq.(\ref{eq:6}).
A similar effects, due to the exciton spin decay, has been observed 
in quantum dots~\cite{QD}.
The present mechanism, instead, is independent on the dimensionality and
driven by the temperature.

From Fig.(\ref{fig:2}) we notice that the B$_2$ and R$_1$ transitions
occur only when a bundle of states get close in energy. The B$_2$ state,
for example, acquires optical strength only when it approaches the
B$_3$ state. {\it The microscopical mechanism of the bright to dark (and vice versa)
transitions is, then, a transfer of optical strength between energetically close
excitonic states}.
In the case of the B$_2$ exciton, for example, this process
occurs by means of a mixing with the B$_3$ state, that
induces an increase of the
contribution from bands with different parity in the e--h pairs embodied in the 
B$_2$ state. This induces a finite dipole, and a finite absorption 
cross--section.
In the case of resonant excitons (with energy larger
than the optical gap $E_g$) this hybridization
is possible because of the continuum e--h substrate which connects the 
states.
However, the bound excitons are discrete states, and the e--h substrate  is replaced
by the energy indetermination due to the finite damping.
This is confirmed by 
a calculation of the {\it h}--BN excitons done
{\it neglecting the exciton damping}, imposing the $E_{\gl}\(T\)$
to be real. In this case the three bound states B$_{1,2,3}$ energies $E_{\gl}\(T\)$
never cross, and the B$_2$  state remains dark at all temperatures.

In conclusion, the electron--phonon coupling induces a severe modification of the 
frozen--atom picture of the excitonic states both at zero and at finite temperatures.
The proposed finite temperature Bethe--Salpeter equation describes, in a fully \ai manner, 
a wealth of new physical
features, and 
makes clear that
a proper and accurate description of the excitonic
states in semiconductors and insulators cannot disregard the coupling 
with the lattice vibrations.

The author thanks X. Gonze, P. Boulanger, and L. Wirtz for fruitful discussions,
and C. Hogan for a critical reading.
I acknowledge support by the European 
Network of Excellence NQ (NMP4-CT-2004-500198).



\begin{thebibliography}{100}
\bibitem{dfpt_calcs}
 The $\frac{\partial{ E_i}}{\partial N}$ factors and the phonon frequencies 
 used to evaluate $\Delta E^{e-p}$ were
 calculated \ai using Density--Functional Perturbation Theory
 in the LDA approximation~\cite{pw}.
\bibitem{kernel_correction}
 The coupling with the nuclear vibrations causes  an additional
 scattering of the $|eh\ra$ pairs with the phonon modes that
 modifies the BS kernel. 
 However this correction is negligible
 for the systems considered in this work.
\end{thebibliography}
\end{document}